\documentstyle[preprint,aps]{revtex}
\tightenlines
\begin{document}
\newcommand{\average}[1] {\left\langle{#1}\right\rangle}

\title{Multiplicity distributions associated to subthreshold events in heavy-ion collisions}

\author{J.~Dias de Deus$^1$, M.~T.~ Pe\~na$^2$, J. C. Seixas$^2$}
\address{$^1$ Instituto Superior T\'ecnico, Dep. F\'{\i}sica - CENTRA, Lisboa, Portugal\\
$^2$ Instituto Superior T\'ecnico, Dep. F\'{\i}sica - CFIF, Lisboa, Portugal\\
}

\maketitle

\begin{abstract}
Subthreshold events (pion production, for instance) at energies $E< m_{\pi}$ are
treated as rare events. The associated multiplicity distribution $P_c (n)$
and the unconstrained distribution $P(n)$, when there is no rare event-trigger, are
related by the model independent relation
$$
P_c (n) = {n^2 \over \average{n^2}} P(n) \ .
$$
This relation,  and in particular its improved version inspired in
clustering of 
nucleons, is in fair agreement with data. Moreover, it allows to extract
from the multiplicity data information on the number of nucleons involved in meson production processes, occuring at energies below the production threshold in free nucleon collisions.
\end{abstract}

\narrowtext
\newpage

Production of particles, pions or other hadrons, energetic photons or fast protons,
in nucleus-nucleus collisions at energies per nucleon well below the
free nucleon-nucleon threshold for such production [1] gives clear evidence for
nucleonic correlations in nuclear matter. These correlations
may simply reflect the fermionic nature of the nucleons or they may
be seen as true collective effects, like clustering of nucleons (see [2] for a review).

By clustering processes we mean here pion production mechanisms 
involving sub-systems with a number of nucleons between 2 and
the mass number A, the simplest example being the interaction of a nucleon with
a deuteron.
In fact, schematically we can represent pion production in the
pd$\rightarrow \pi^0$pd reaction via resonance formation as
in the diagram
of Fig.1a. The threshold for the mentioned reaction
is lower than the free nucleon threshold (Fig.1b).
In contrast to this diagram, the one of Fig.1a is kinematically allowed at subthreshold
production energies, because the
two nucleons of the deuteron are correlated.  
Basic diagrams [3-6], like the one
in Fig.1b, require a medium
to be non vanishing, for energies below the production energy threshold.

Higher mass resonances can be obtained if larger clusters supply the required
energy. Our emphasis here is on clustering rather than on the specific mechanism
of pion production (like higher-mass resonances [7,8])

In order to see, in a simplified manner, what the problem is,
let us write the laboratory energy $E_L$ of nucleus $A$ as
$$
E_L= A \average{E}_N \ ,\eqno(1)
$$
where $\average{E}_N$ is the average nucleon energy.
If one assumes that a collision results from
the superposition of nucleon-nucleon collisions and ignores Fermi momentum and binding energy, the threshold kinetic energy per nucleon to produce, say, a pion is, 

$$
E_{th} \equiv \average{E}_N - m_N |_{th} = 2m_N \left[ (1+{m_{\pi} \over 2m_N})^2 -1\right] \simeq 2m_{\pi} \simeq 280\ {\rm MeV} \ . \eqno(2)
$$
This is the free nucleon threshold. In nuclear matter
one may have nucleons with energy above $\average{E}_N$ and the threshold
becomes lower than (2).

The lowering of the threshold (2)
can be very easily visualized if one allows
for clustering of nucleons.
If $\alpha \leq A$ nucleons of nucleus $A$ --- with mass number A--- collide
with one nucleon of nucleus $B$ --- with mass number B --- (or vice-versa)
the threshold energy per nucleon becomes,
$$
E_{th} \simeq {1 +\alpha\over \alpha} m_{\pi} \geq m_{\pi} \simeq 140\ {\rm MeV} \ . \eqno(3)
$$
This is, of course, the threshold for free nucleon-nucleus collisions.

As experimental production of pions occurs even at energies bellow $m_{\pi}$,
this requires clustering from both nuclei,
$\alpha \leq A$ and $\beta \leq B$, 
$$
E_{th} \simeq {\alpha +\beta \over \alpha \beta} m_{\pi} \ ,\eqno(4)
$$
with the absolute threshold, naturally occuring for $\alpha =A$ and $\beta=B$,
$$
E_{th} \simeq {A +B \over AB} m_{\pi} \ .\eqno(5)
$$

In the case of $^{12}C-$$^{12}C$ interactions, for instance, this threshold
corresponds to $\simeq 23$ MeV. The binding energies will affect this
value within less than a keV, provided  they are considered simultaneously
in the final and initial states. This lower bound cannot be obtained in a simple way from Fermi-motion based arguments.

Independently of the underlying model for the nucleus-nucleus collisions,
the point we would like to make is that these sub NN threshold events
are {\bf rare}, in the sense that their probability of occurence is very small.
While total inelastic cross-sections are of the order of several
milibarn, the cross-sections we are talking about here are of the order of
the microbarn or nanobarn. In other words, by imposing kinematic restrictions
through lowering the energy available in the system,
one moves to the tails of the fermionic distributions or requires simultaneous
clustering, and the events become rare, the probability of occurence being
very small.

It is interesting to note that as one unconstrains the kinematics, i.e.,
increases the energy, the $\pi$ production cross-section increases very rapidly
reaching the milibarn values for $E_{th} > 2m_{\pi}$ (see Fig.(2)).
That is the region of (free) nucleon-nucleon interactions where
the Glauber approach becomes valid.

We shall next make a short discussion on rare events (see references [9]),
adapted to the present situation, where we are interested in clustering
processes in {\it both} colliding nuclei.
If in a multi nucleon-nucleon collision process,
$\nu$ elementary collisions occur, and $\tau_c$ is the probability
of clustering nucleons, and $N(\nu)$ is the number of events with
$\nu$ elementary collisions, we can write the identity 
$$
N(\nu) = \left[ \Sigma^{\nu}_{k=0} \pmatrix{\nu \cr k} \tau_c^k (1-\tau_c)^{\nu -k}\right] 
\left[ \Sigma^{\nu}_{k'=0} \pmatrix{\nu \cr k'} \tau_c^{k'} (1-\tau_c)^{\nu -k'}\right] N(\nu) \ .\eqno(6)
$$
The meaning of (6) can be easily grasped by expanding in
powers of $\tau_c$ to obtain,
$$
N(\nu) = \left[ (1-\tau_c \nu)^2 + 2(1-\tau_c \nu) (\tau_c \nu) +
(\tau_c \nu)^2 + \ldots \right] N(\nu) \ .\eqno(7)
$$
In this sum the first term means the normal contribution
(Glauber-type, i.e. no clustering in both nuclei, giving \mbox{$E_{th} \simeq 2 m_{\pi}$}),
the second term means cluster in one of the nuclei (reducing $ E_{th}$ to
\mbox{$ E_{th} \simeq m_{\pi}$}) and the third term,
$$
N_c (\nu)= \tau_c^2 \nu^2 N(\nu) \ ,\eqno(8)
$$
represents what interests us, namely nucleon clustering in both nuclei (allowing $(E_{th} < m_{\pi}$). The subsequent contributions, corresponding to multiple clustering are negligeable. In a sense, the events described by (8) are doubly-rare events.

If one makes the reasonable assumption that the number $\nu$ of collisions is a measure of the number $n$ of produced particles, from (8) we obtain the $\tau_c$ independent, universal relation
$$
P_c (n) = {n^2 \over \average{n^2}} P(n) \ , \eqno(9)
$$
where $P(n)$ is the unconstrained particle multiplicity distribution, $\average{\ldots}$ denotes the average value with respect to the $P(n)$ distribution and 
$P_c (n)$ is the multiplicity distribution associated to the rare event ($\pi$ emission below the proton-nucleus threshold). 

Note that
$$
\Sigma_{n=0} P_c (n) = \Sigma_{n=0} P(n) =1 \ ,\eqno(10)
$$
and that [9] 
$$
\average{n}_c = {C_3 \over C_2}\average{n} \ ,\eqno(11)
$$
where $C_q \equiv \average{n^q}/\average{n}^q$. Equation (11) implies that
$$
\average{n}_c    >     \average{n} \ . \eqno(12)
$$
It is also clear that 

$$
P_c (n)  < P(n)\  \mbox{$\Leftarrow$} \ n^2 < \average{n^2} \ , 
$$
$$
P_c (n) > P(n)\ \mbox{$\Leftarrow$} \ n^2 > \average{n^2} \ , \eqno(13) 
$$
i.e., the distributions must cross at some point $n$, corresponding to
{$n^2 = \average{n^2}$}.

In Fig.\,3 we test Eq.(9) for the particular case of the data of ref. [10] for
$^{36}$Ar on $^{27}$Al collisons.
The data $P(n)$ [10] on  the inclusive charged particles
distribution (nucleons, etc...) is shown by the open circles.
The data $P_c (n)$ [10] on the equivalent distribution when a sub-threshold pion,
at $E=95$ MeV/Nucleon, is detected is shown by the open squares.
Qualitatively, equation (9), implying equations (11) and (12), satisfies the data: 
the average
multiplicity, with a trigger on $\pi$, is larger, $\average{n}_c \simeq 9$
while $\average{n} \simeq 4.4$ [11], the two
distributions cross at $n^2 = \average{n^2}\simeq 49 > \average{n}^2 \simeq 19.4$
(the inequality $\average{n^2} > \average{n}^2$ has to be satisfied), the
distribution is independent of the rare event trigger [12].

For the description of $P(n)$ (full line) the generalized gamma function
was used

$$
P(n)={1\over \average{n}} {\mu \over \Gamma (k)} \left[ {\Gamma (k+1/\mu)\over \Gamma (k)}\right]^{k\mu} \ \left( {n\over \average{n}}\right)^{k\mu-1}
$$

$$ 
\times \exp \left[ -\left( 
{\Gamma (k+1/\mu) \over \Gamma (k)} {n\over \average{n}}\right)^{\mu} \right] \eqno(14)
$$
with parameters
$k= 0.77$, $\mu =2.0$, $\average{n} =4.4$. To describe $P_c (n)$  equation
(9) was used. The results (short-dashed line) show that
the agreement of (9) with the rare event triggered
distribution is not quantitative enough, namely to allow a correct
description of the crossing point with the inclusive distribution
curve.

To improve this situation, one should notice that in the case of the production
of a pion, subsequent absorption may affect
the relation (9). The multiplicity associated to a fast photon, for
instance, would be better to test (9) as the photon is less affected
by absorption.

On the other hand, recently,
in the
context of nucleus-nucleus collisions
at high energy it was argued that when clustering
effects occur the multiplicity has an additional increase because
the clusters themselves produce more particles [13]. In other words,
(9) has to be modified to become
$$
P_c (n+\delta) = {n^2 \over \average{n^2}} P(n) \ ,\eqno(15)
$$
where $\delta$ traces the  average number $\alpha$ ($>1$) of nucleons in
the clusters, accordingly to
$$
\delta \simeq 2 \alpha -2 \ .\eqno(16)
$$
For $\alpha=1$, $\delta=0$ as it should (no clustering) and (15) reduces to (9).

Using (4), with $\alpha \simeq \beta$, and requiring
$$
E_{th} = {2\over \alpha} m_{\pi} >  95\ {\rm MeV/Nucleon} \eqno(17)
$$
one estimates
$$
\alpha \leq 3 \eqno(18)
$$
and, from (16),
$$
\delta \leq 4 \ .\eqno(19)
$$

In Fig. 3 we also included the description
of the pion-triggered data
$P_c(n)$ by means of equation (15) with
$\delta =3$ (long-dashed line). In comparison with
the  $P_c(n)$ results given by (9) (short-dashed line),
improvement is achieved with the curve corresponding to (15).
Indeed the shift by $\delta$ originated by cluster
formation describes better
the cross-over point of
the constrained and un-constrained multiplicities. Moreover, 
the value of the parameter $\delta$, which is found to be consistent
with the distribution data, is also consistent with (18) and (19), originated
only from the observed sub-threshold energy.
Thus, Eq.\, (19) constitutes an important bound constraining the overall behavior of the data. In conclusion, clustering arguments relate directly
the observed sub-threshold production energies with the behavior of the
distribution data.

The physical picture is that at sub-threshold energies the corresponding
wavelength scale of the process is large enough for
clusters of nucleons in the nucleus to recoil
as a whole. At these energies correlations due to nucleon-nucleon interations
may dominate Pauli blocking or Fermi-motion correlations.
An absolute threshold may be calculated corresponding to
a recoil of the complete nucleus as a whole (equation (5)).
The deviation of the physical
threshold from the absolute one is due to clustering formation involving
only part of the constituent nucleons. The clustering also affects the distribution 
data for particle production events triggered by rare events such as pion production.

Therefore, rare-event triggered measurements give direct information
on the number of nucleons involved in particle production at
threshold. Accordingly, few-nucleon production reactions, near threshold,
such as $pd \to {\pi^{\circ}}\,$$^3$He and $pd \to \eta\,^3$He for example,
which may be exactly calculable at present, 
are worth being investigated experimentally at the 
existing strong focusing synchrotons facilities (COSY, CELSIUS), since
they may confirm the proposed relation
between threshold energies and number of nucleons involved in the
production mechanisms.
We note here also that the failure already found in reference
[14] to describe by means
of the impulse approximation the $^3He(\pi^¯,\eta)^3H$ reaction 
below the free production threshold for forward scattering,
is already an evidence for a cluster-enhanced rare event.

As we mentioned before, it is not our purpose to test models of
interacting nuclei but to test the validity of (9) when the kinematic
constraints are such that occurence of event c (pion, fast proton,\ldots)
is rare. However, Eq.(15), which in any case is approximate, as $\alpha$
and $\beta$ may fluctuate, only makes sense in a model with interaction via
clusters. This is similar to what is proposed in [13] for cumulative effects at very
high energies. Models with independent $NN$ collisions ($\alpha=1$)
in medium (based on Fermi-motion distributions arguments alone) do not give
support to the arguments leading to (15). So, although relation (9) is universal, (15) depends upon
assumptions concerning the type of interations involved and thus can distinguish between models.

The fact that particles can be produced {\em below threshold} via a clusterization
mechanism raises still another important question, namely, whether particle production
normally forbidden by Zweig's rule could occur in an environment where (microscopic)
clusterization takes place. In particular, this effect 
may have important consequences for $\phi$ meson production in heavy ion collisions at
SPS and LHC energies. It remains also to be clarified whether this effect has any consequences for $J/\psi$ production at collider energies.

\bigskip
\vspace{3cm}
\noindent
{\bf Acknowledgements}\\
\noindent\medskip

The authors want to thank S. A. Coon for helpful discussions, as well as W. Cassing and C. Louren\c co for their comments.

\newpage
\hspace{5cm} Figure Captions\\

\vspace{1.5cm}
\noindent
Figure 1 - Pion production mechanisms mediated by resonance formation 
in the medium (a) and in free nucleon scattering (b) .\\

\noindent
Figure 2 - Experimental cross-section for $^{12}C$-$^{12}C$ from the fourth
reference in [1]. The full line is only to guide the eye.\\

\noindent
Figure 3 - Data for $^{36}Ar$ on $^{27}Al$ collisions taken from reference [10].
The full curve is the fit to $P(n)$,
the short-dashed line is $P_c(n)$ given by equation 9,
the long-dashed line is $P_c(n)$ given by equation 15.


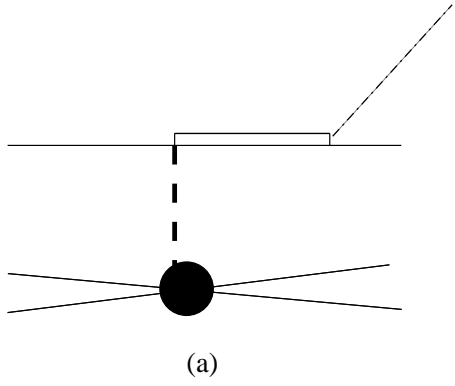
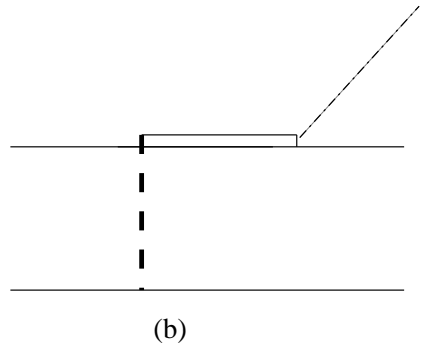
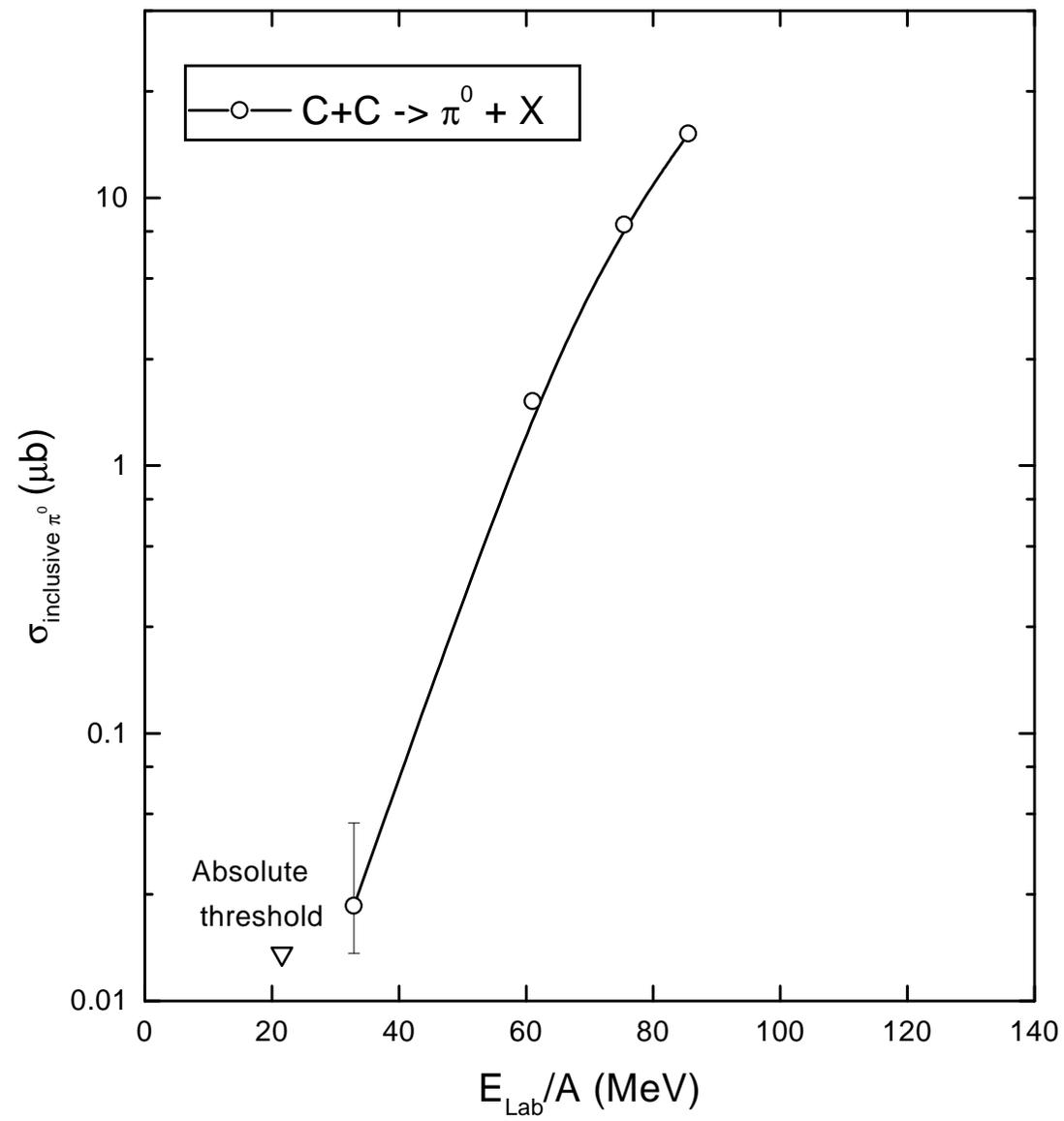
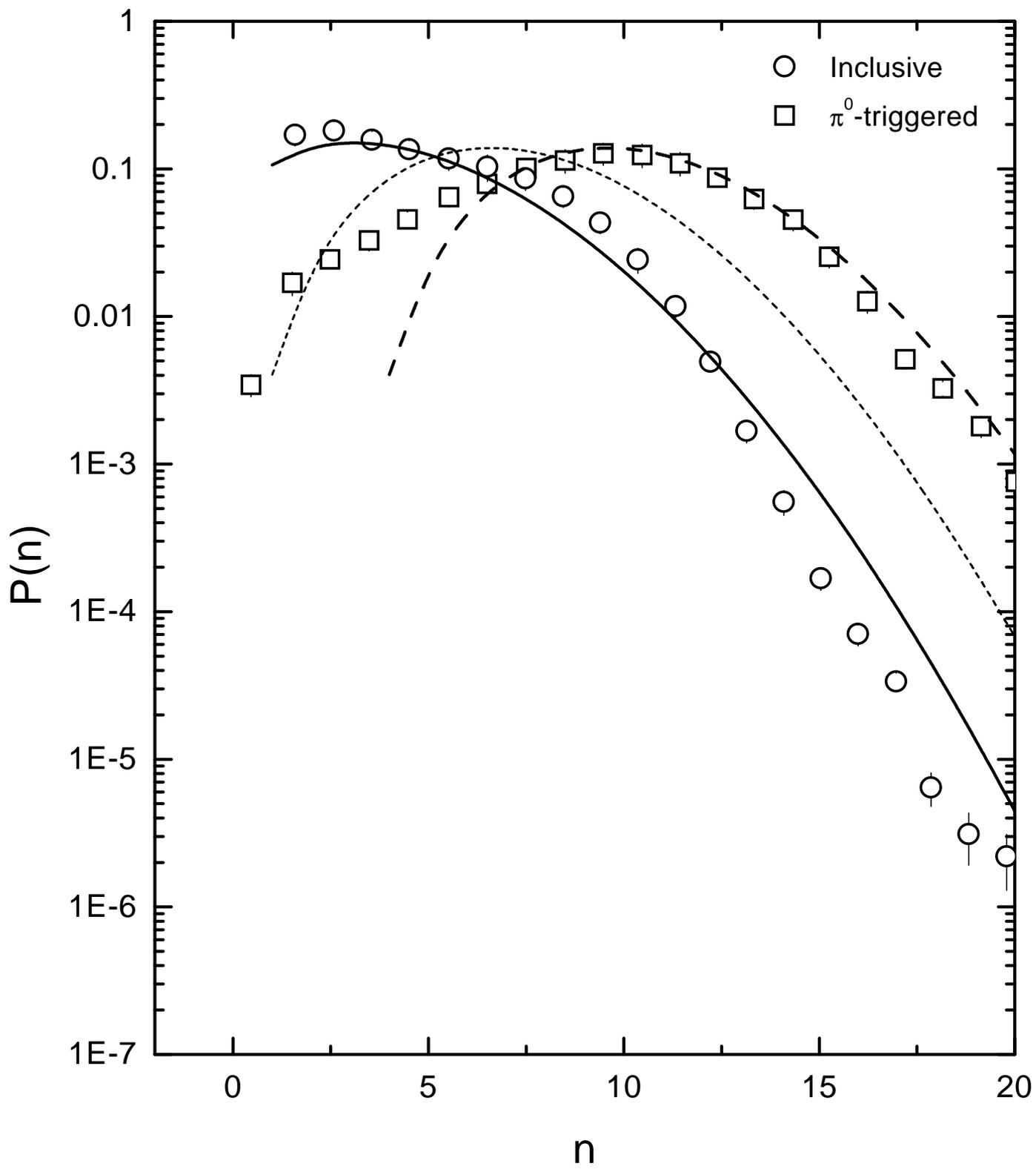
\end{document}